\documentclass[12pt]{article}
\usepackage[margin=1in]{geometry}
\usepackage{paralist}
\usepackage{longtable}
\usepackage{array}
\usepackage{setspace}
\usepackage[title]{appendix}
\usepackage{xurl}
\usepackage[table]{xcolor}
\usepackage[colorlinks=true, urlcolor=blue]{hyperref}
\RequirePackage{amsthm,amsmath,amsfonts,amssymb}
\RequirePackage{algorithm,algorithmic}
\usepackage{thmtools}
\RequirePackage{xcolor}
\RequirePackage{bbm}
\RequirePackage{textcomp}
\RequirePackage{gensymb}
\RequirePackage{centernot}
\RequirePackage{bm}
\RequirePackage{enumerate}
\RequirePackage{enumitem}
\RequirePackage{graphicx}
\RequirePackage{multirow}
\RequirePackage{booktabs}
\RequirePackage{mathtools}

\theoremstyle{plain}

\newtheorem*{theorem*}{Theorem}

\newtheorem*{claim*}{Claim}

\newtheorem*{lemma*}{Lemma}

\newtheorem*{proposition*}{Proposition}

\theoremstyle{remark}

\renewcommand\thmcontinues[1]{Continued}

\usepackage{crossreftools}
\makeatletter
\newcommand{\optionaldesc}[2]{%
  \phantomsection
  #1\protected@edef\@currentlabel{#1}\label{#2}%
}
\makeatother

\usepackage{hyperref}
\usepackage{caption} 

\newcommand{\adj}{\textnormal{adj}}

\title{Evaluating the Impact of Rhode Island’s Self-Sustaining Reemployment Services and Eligibility Assessment (RESEA) Program on Employment Outcomes}
\author{Harrison H Li\thanks{Department of Mathematics, Harvey Mudd College. \url{harrli@hmc.edu}} \\
Shanna Pearson-Merkowitz\thanks{School of Public Policy, University of Maryland, College Park} \\
David Yokum\thanks{School of Data Science and Society, University of North Carolina at Chapel Hill}
}
\date{}

\makeatletter
\renewcommand{\@biblabel}[1]{}
\makeatother
\begin{document}
\maketitle

\begin{abstract}
\begin{singlespace}
Prolonged unemployment carries serious economic, health, and wellbeing costs. With federal support, most U.S. states now operate a Reemployment Services and Eligibility Assessment (RESEA) program to help Unemployment Insurance (UI) claimants return to work faster. We report results from a large (N = 23,549) preregistered randomized controlled trial (RCT) evaluating Rhode Island's RESEA program from February 2022 to September 2023. We estimate that selection into the program increased annualized wages by \$1,153, increased reemployment by 1.5 percentage points, and reduced UI duration by nearly two weeks. The vast majority of these wage and reemployment effects appeared within two quarters of claimants’ first pay dates and persisted through at least the following year, and we estimate that each dollar spent on the program saved the state \$2.64. Using causal forests, a machine learning technique for estimating heterogeneous treatment effects (HTE), we also conduct an exploratory analysis to investigate if there are differential effects of selection into the RESEA program. We find that all participants experienced positive wage benefits from RESEA selection, with particularly large effects for older and lower-income workers. Finally, we improve upon prior RESEA evaluations by explicitly controlling for the week of treatment assignment — a methodological refinement absent from several existing RCTs of job-training programs that is important to eliminate confounding bias. We also discuss ways to harvest precision gains from baseline covariate adjustment without introducing large-sample bias. 
\end{singlespace}
\end{abstract}

\section{Introduction and Background}

The Social Security Act (SSA) of 1935 established unemployment insurance (UI) programs in the United States, ensuring that many workers who involuntarily lose their jobs can receive a temporary wage replacement until they are reemployed. While partial wage replacement serves as a critical safety net to help ensure financial stability while the recipient identifies and secures a new job, there are longstanding concerns that it could also prolong unemployment by decreasing the incentive to get back to work quickly. While there may be benefits to taking time to get back to work, UI programs also must manage their finances and ensure there are sufficient funds for all those who are or may become unemployed. A long line of research also suggests that prolonged unemployment not only is detrimental to the short-term finances of workers, but also carries considerable long-term costs, including increases in morbidity and other mental and physical health challenges as well as wage decreases that carry over to the next generation (e.g., Nichols et al. 2013; Shiro \& Butcher 2022). 

U.S. states began experimenting in the 1990s with various interventions aimed at helping their unemployed workers reenter the workforce quickly. These measures included enforcing job search requirements, providing job training assistance, and offering cash bonuses for rapid reemployment. Each of these was shown to reduce UI duration in randomized controlled trials (Corson et al., 1985; Meyer, 1995; Corson et al., 1989; Anderson et al., 1991; Corson \& Haimson, 1996; Hanna \& Turney, 1990). Many (but not all) of these studies also showed improvements to claimants’ wage and employment outcomes. This work encouraged Congress to establish the Worker Profiling and Reemployment Services (WPRS) program in 1993, which mandated targeted reemployment services to UI claimants with high predicted probabilities of exhausting their UI benefits based on factors like claimant demographic information, industry, and prior earnings (Black et al., 2002).

Today, most states incorporate WPRS’s targeting mandate into Reemployment Services and Eligibility Assessment (RESEA) programs (O’Leary \& Pepin, 2021, Trutko et al., 2022)  and in 2025, the U.S. government allocated \$345,320,000 in grants for the RESEA program (DOL-ETA, 2025). RESEA programs are a successor to earlier Reemployment and Eligibility Assessment (REA) programs, which formalized one-on-one eligibility verification in addition to job seeking assistance. Starting in FY 2023, states were required to allocate at least 25\% of grant funds specifically to interventions that had already demonstrated strong causal evidence for improving outcomes or be actively evaluating their program through an experimental design that provides strong causal evidence. 

In this paper, we provide a contemporary causal evaluation of the RESEA program in Rhode Island. Specifically, our primary analysis estimates the program’s average intent-to-treat (ITT) effect on three preregistered outcomes relating to post-treatment wages, UI duration, and reemployment. We also report short term effects (second quarter after treatment) to evaluate when the effects take place and how long they last. The broader policy objective is to rigorously assess whether the RESEA model employed in Rhode Island served workers under macroeconomic and demographic conditions that differed substantially from those in the original studies that motivated federal adoption of RESEA. As a secondary analysis, we use statistical methods not previously applied in this context to examine which participants benefitted most from selection into RESEA.

\subsection{Literature Review}

The evaluation literature on REA, RESEA, and similar predecessor programs suggests that these programs can effectively accelerate reemployment and decrease the duration of unemployment insurance (UI). Across multiple state evaluations, researchers have frequently found that treatment groups experience statistically significant reductions in total weeks spent on UI and the overall amount of UI benefits received. For instance, early evaluations of the New Jersey Unemployment Reemployment Demonstration Project observed meaningful drops in both UI benefit dollars and weeks collected from job search requirements and/or cash bonuses for early reemployment (Anderson et al., 1991; Corson \& Haimson, 1996). A randomized study in 1986-87 in Washington similarly found that removing work-search requirements and work tests led to higher rates of UI benefit exhaustion and lower subsequent employment (Lachowska et al., 2015; Lachowska et al., 2016). More recently, a set of RCTs of REA in several states showed that the program reduced the probability of benefit exhaustion while simultaneously boosting short-term and long-term earnings and employment rates (Manoli et al., 2018; Michaelides \& Mian, 2021; Michaelides et al., 2012; Poe-Yamagata et al., 2011).

While the overall body of evidence highlights program success, there are nuances based on geographic location, program intensity, and specific operational designs. In some contexts, such as an evaulation in North Dakota, the REA treatment yielded no statistically significant impacts on benefit receipt, employment, or earnings (Benus et al., 2008). Similarly, a multi-state Job Search Assistance Demonstration found that while treatment successfully reduced UI benefits and increased earnings in Washington, D.C., it produced no such impacts in Florida (Decker et al., 2000). Recently, some research has shifted to evaluating not just if these programs work overall, but if specific elements are effective. For example, research in Indiana, New York, Washington, and Wisconsin suggests that claimants mandated to participate in multiple sessions spend fewer weeks on UI than those receiving only a single intervention (Klerman et al., 2019).

Methodologically, we group related prior work into three categories. In one group are surveys and other observational studies, which cannot reliably separate causation and association due to the possibility of unobserved confounding. We divide the experimental studies, which have dominated the research on RESEA and related programs, into two categories based on their design: simple random assignment (e.g. Benus et al., 2008; Decker et al., 2000; Lachowska et al., 2015; Lachowska et al., 2016) and block randomization (Anderson et al., 1991; Brigandi et al., 2024; Corson \& Haimson, 1996; Corson et al., 1989; Klerman et al., 2019; Manoli et al., 2018; Michaelides \& Mian, 2021; Michaelides \& Mueser, 2018; Michaelides et al., 2012; Poe-Yamagata et al., 2011). With either block randomization with equal block sizes and equal fractions receiving treatment across all blocks (Anderson et al., 1991; Corson \& Haimson, 1996; Corson et al., 1989; Klerman et al., 2019) or simple random assignment, differences in mean outcomes (possibly with regression adjustment for observed baseline covariates) or linear regression with fixed effects for the blocks can give asymptotically unbiased estimates for the average treatment effect. However, block randomization is widely utilized across different local offices or time horizons, and external constraints like the number of UI applicants and/or staffing constraints often make the treatment fractions across blocks unequal. Then block identity is an important confounder that needs to be appropriately adjusted for to avoid bias. We discuss this point further in the Discussion section of this manuscript. In Appendix Table 1, we provide a full review of the evaluation literature that is included in the U.S. Department of Labor’s Clearinghouse of Labor Evaluation and Research (CLEAR) database for studies of RESEA with high causal evidence. The table summarizes each study’s findings and methodological considerations. We note that none of these studies directly evaluate RESEA, but they are all deemed by the U.S. Department of Labor (USDOL) to focus on relevant similar job training programs for UI recipients.

\section{Methods}
\subsection{Outcome definitions}
Our preregistered wage outcome is defined to mirror the computation of base wages performed by the Rhode Island Department of Labor and Training (RIDLT) to determine weekly UI benefit rates and credit balances.\footnote{A copy of our pre-analysis plan is available at \url{https://osf.io/c9xe8/}.} It is defined as the sum of the two highest quarterly wages for each claimant among the second through fifth complete calendar quarters (inclusive) following the randomization date, multiplied by two for annualization, and ranges from \$0 to \$630,875.\footnote{We additionally preregistered an evaluation of the complier average causal effect (CACE) that showed larger effect sizes with a similar level of statistical confidence as our main results. We decided ex-post to focus on the ITT effect here due to the potential pitfalls of the instrumental variable assumptions of the CACE analysis.} Our reemployment outcome is a binary variable defined as 1 if at least two of these quarters had positive (nonzero) reported wages. Finally, the number of weeks of UI benefits drawn were estimated based on credit balances and weekly benefit rates in the 78 weeks following the claim effective date, and range from 0 weeks to 51.1 weeks. The wage and reemployment outcomes were computed using RIDLT’s wage records. These records are produced quarterly and include all formal wages paid and reported by Rhode Island employers. The records do not capture self-employment income or income earned out of state. As a result, our wage and reemployment estimates may be biased downward because claimants who lost covered employment but subsequently become self-employed or found work out of state are recorded as having zero earnings and thus as not reemployed. We address this and conduct exploratory analysis on this issue below.

We also analyze and present “short-term” variants of these three outcomes, following the CLEAR guidelines established by the USDOL for RESEA intervention effectiveness (CLEAR 2015/2022). These outcomes are the reported wage in the second complete calendar quarter after the randomization date (multiplied by 4 for annualization), whether this reported wage was nonzero, and duration of UI in the same benefit year as the claim (out of a maximum of 26 weeks).

\subsection{The RESEA program and study design}

On each Wednesday from February 9, 2022 to September 27, 2023, inclusive (hereafter the study period), approximately 150 (later 135) newly eligible UI applicants in Rhode Island (those whose first UI payments were scheduled for that week) were randomly selected for RESEA (treatment group). Those selected for RESEA received a letter informing them they were required to make a meeting appointment within 18 days of receiving the selection letter. If the participant did not either make an appointment and meet with a job coach or report a return to work date in this time, their case was referred to adjudication and their benefits may have been suspended. The claimants not selected for RESEA each week (control group) did not receive a letter and were not required to schedule or attend a meeting. While control group claimants still had voluntary access to RESEA services, only 5 such claimants opted in during the study period. 

The program included the following components. Once an initial RESEA meeting was scheduled, the claimant received a packet of information about RESEA. The packet included compliance requirements, information on UI fraud, a list of services and contact information, information about job training opportunities, extensive tips for interviewing, and a job search form in which the claimant was to keep track of their job search activities which RIDLT uses for compliance. The counselor meetings were primarily held virtually in response to the COVID-19 pandemic—a significant departure from previous programs—though claimants could also attend appointments in-person at local career centers on a first come, first served basis. During each RESEA meeting, the RESEA counselor would confirm UI eligibility and the participant's ability to work. This was followed by a skills evaluation and job search guidance. Finally, they would review labor market information and discuss career goals. The session included guidance on using the state’s website for job searches, resume building tailored to job specifications, and, if deemed necessary or desired, a mock interview. Typically if the claimant was laid off from a declining industry, participants skills assessment were used to help identify career opportunities in sectors with more job openings. The program also provided referrals to other programs that help support job reentry and social services when appropriate. Finally, RESEA participants were required to submit a 30-day work search report at a second meeting.

Across all weeks of the study period, there were a total of 11,700 UI claimants in the treatment group and 11,849 claimants in the control group.\footnote{ This study was determined by the Brown University Human Research Protection Program (HRPP) Office of Research Integrity (ORI) Office that it did not require IRB approval on May 18, 2020 as it was considered "program evaluation." Participant data was de-identified and part of a data sharing agreement for ongoing projects of the Policy Lab at Brown University and the State of Rhode Island. } In sum, we ran a randomized block experiment where the blocks correspond to weeks. We employed the following algorithm for randomizing treatment assignments each week. First, all eligible claimants noting veteran status were deterministically assigned an integer value of 9,999,999,999 due to regulatory requirements that they were always selected for RESEA. Then, all remaining eligible claimants were randomly assigned an integer between 0 and 9,999,999,999, inclusive. Initially, the 150 claimants assigned the highest integers each week were to be selected for the treatment group. This number was reduced to 135 starting on May 2, 2023 due to small control group sizes resulting from fewer than expected UI applicants, as well as staffing limitations. This decision was made before any of the outcome data was examined, and thus does not hinder the validity of our statistical conclusions.

The random integer generation for the non-veteran claimants was designed to be uniformly random: each applicant’s social security number was added to the six-digit microsecond timestamp at the time of assignment, and then the digits were written out in reverse to form the seed of a uniform random integer generator in COBOL. To evaluate whether this randomization worked properly, we performed weekly balance checks. Specifically, we computed p-values from two-sample permutation tests on the data from each week to check whether various pre-treatment covariates were balanced between the treatment and control groups among those in the study cohort (i.e., excluding veterans and claimants whose claims were later deemed fraudulent). These permutation tests are nonparametric and based on a test statistic $T$ defined as the absolute value of the difference in mean covariate value between the two groups (binary covariates were encoded as 0 or 1). To obtain $p$-values, we randomly shuffled the group labels $M = 10,000$ times, computing test statistics $T_i$ on each such permuted dataset $i = 1,\ldots,M$. Finally, we computed a two-sided $p$-value based on the number of test statistics $T_i$ larger than the test statistic $T$:
\[
p=\frac{1+\#(T_i >T)}{1+M} 
\]

Under the null hypothesis that the control and treatment distributions of the covariate are exchangeable, this $p$-value is stochastically larger than the uniform distribution on $[0,1]$. In Fig. 1, we observe this appears to be the case for all of the covariates considered.

\begin{figure}[!tbp]
\centering
\includegraphics[width=\linewidth]{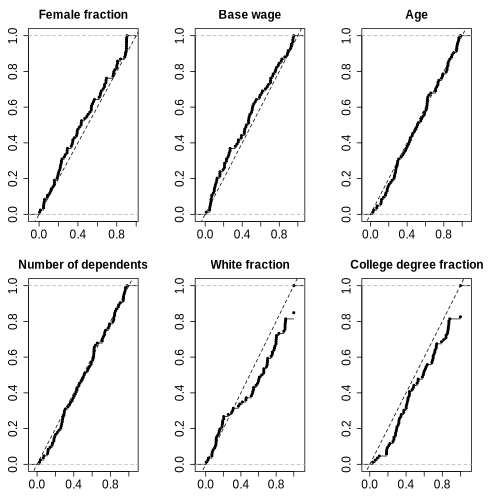}
\caption{For selected covariates, the empirical cumulative distribution function (ECDF) of the p-values of the weekly balance tests across all weeks during the study period. Data consistent with the null hypothesis of proper treatment randomization would have an ECDF near or below the dashed diagonal line.}
\label{fig:balance}
\end{figure}

\subsection{Data filtering}
\begin{figure}[!tbp]
\centering
\includegraphics[width=\linewidth]{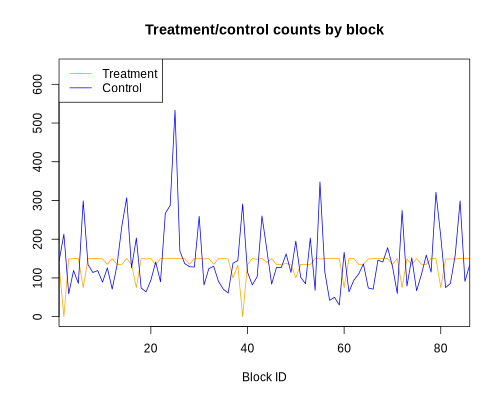}
\caption{The number of individuals assigned into the treatment (i.e., selected for RESEA) and control (i.e., not selected for RESEA) groups in each block of our randomized controlled trial. Note that the block ID’s were randomly assigned to preserve anonymity, and that these counts include all claimants (total n = 24,120), including those excluded from our study cohort (veterans and those whose claims were later determined to be fraudulent)}
\label{fig:weekly_trt_ctl}
\end{figure}

There were a total of 24,120 individuals who successfully applied for UI in Rhode Island during the study period. The number of individuals selected for the treatment group did not exactly equal 150 (or 135) in every week due to issues such as technical errors and fluctuating staffing availability for appointments (Fig. 2). To preserve the causal fidelity of our conclusions, which are based on treatment randomization, we exclude from our analysis all 502 claimants from the two weeks during the study period on which nobody was randomized due to such issues. We additionally exclude from our analysis the 27 individuals whose claims were later determined to be fraudulent as well as all 42 veterans because including the veterans would undermine the causal validity of our study as all veterans are selected for RESEA for legal reasons, as noted above. This leaves $n = 23,549$ individuals in our final study cohort. There was also no matching credit balance information for 22 of the claimants in the study cohort, so we were unable to estimate the weeks spent on UI for these claimants, and additionally exclude them from all of our analyses pertaining to the time spent on UI.

\subsection{Compliance}

\begin{figure}[!tbp]
\centering
\includegraphics[width=\linewidth]{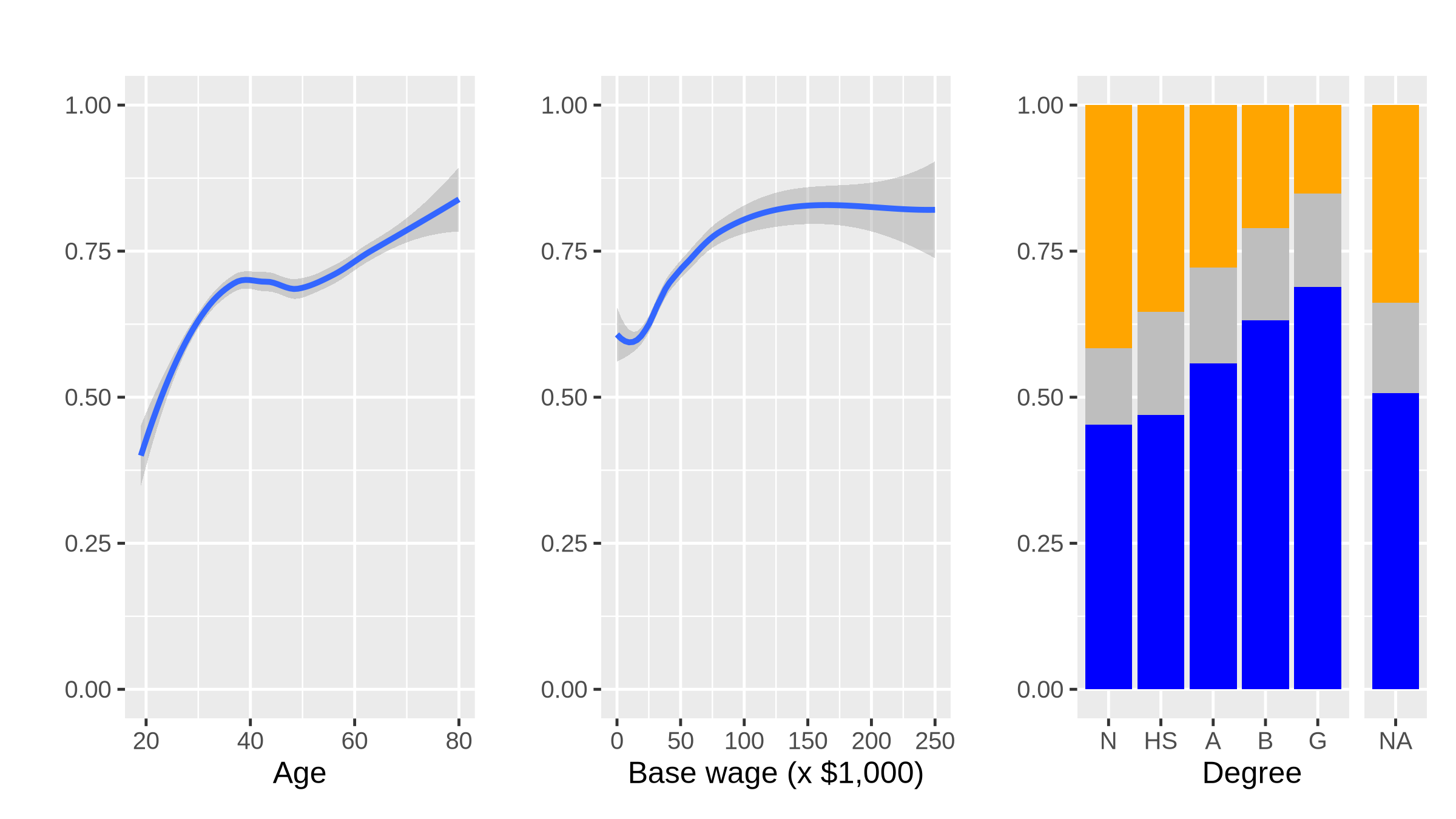}
\caption{Smoothed RESEA compliance rates among the treatment group as a function of age (left, truncated at 80 years) and base wage (center, truncated at \$250,000), as well as a breakdown of RESEA completion status by highest degree obtained (right). The smooths (blue) are computed using locally weighted regression (\texttt{loess} in R with \texttt{span} = 0.7). Note: In the right panel, the blue bars correspond to individuals who successfully completed RESEA, the grey bars correspond to those who were found to be exempt from completing RESEA after selection (most commonly due to already having a return-to-work date), and the orange bars correspond to individuals who did not successfully complete RESEA. For the left and center panels, exempt individuals were viewed as compliant. Degree abbreviations: N = No high school diploma, HS = High school diploma or GED, A = Associate’s degree or some college, B = Bachelor’s degree, G = Graduate degree, NA = Education information not available (phone filer)}
\label{fig:compliance}
\end{figure}

Although RESEA compliance is enforced by the threat of termination of UI benefits, around 31 percent of claimants in the treatment group (selected for RESEA) did not attend the required meeting. Higher rates of not attending the meeting were observed among those who were younger adults, lower earning, and less educated (Fig. 3). Since receiving the letter appears to often result in claimants reporting a return to work date to remain in compliance while not attending RESEA services, there is likely a direct causal effect from selection into RESEA, independent of program completion. In particular, receiving a letter requiring a mandatory action to continue receiving benefits, which may motivate claimants to expedite their job search without attending the required meeting or to report information they would have eventually reported but later that removes them from the RESEA pool. Such an effect has been proposed in previous work (Black, Smith, Berger, \& Noe, 2002) and directly invalidates the assumption, necessary for identifying the complier average treatment effect (CACE) in a randomized controlled trial, that selection into RESEA is an instrumental variable for completion of the program (Imbens \& Angrist, 1994).

An alternative method to estimate the difference between the causal effect of completing RESEA and the effect of merely being selected for RESEA would be to attempt to adjust for any differences in outcomes between RESEA compliers and non-compliers based on observed individual covariates, e.g., via matching or reweighting methods. However, the limited number of demographic variables available in our dataset makes it difficult to be confident in the assumptions of such methods, which would require the absence of unobserved confounders (individual attributes associated with both compliance and the outcome) and thus not meet CLEAR’s standards for high strength of causal evidence.

\subsection{Primary analysis methodology}
We preregistered block size-weighted averages of difference-in-means estimators as our estimates of the ITT effect of RESEA on our outcomes of interest. Specifically, let $Y_{ti}$ be the outcome of interest (for instance, our wage, reemployment, and weeks on UI outcomes, or their short-term counterparts, as defined above) for subject $i=1,\ldots,n_t$ in block $t=1,\ldots,T$ and let $W_{ti}$ be the binary treatment indicator for that subject (1 if selected for RESEA, 0 if not selected). Then our point estimate of the ITT effect for the outcome is
\begin{equation}
\label{eq:tau_hat}
\hat{\tau} = \sum_{t=1}^T \frac{n_t}{n}\hat{\tau}_t, \qquad \hat{\tau}_t = \frac{\sum_{i=1}^{n_t}W_{ti}Y_{ti}}{\sum_{i=1}^{n_t} W_{ti}} - \frac{\sum_{i=1}^{n_t}(1-W_{ti})Y_{ti}}{\sum_{i=1}^{n_t}(1-W_{ti})}
\end{equation}

These computations are implemented by the \texttt{difference\_in\_means} function in the \texttt{estimatr} package in the R programming language (Blair et al., 2025), which also provides associated asymptotic confidence intervals and $p$-values based on a central limit theorem approximation. Since our block sizes are fairly large (in the hundreds), this approximation should be quite reasonable. 

In an attempt to further improve precision, we also report covariate-adjusted ITT effect estimates $\hat{\tau}_{\adj}$. These estimates are, like $\hat{\tau}$, also block-size weighted averages of within-block estimators:
\begin{equation}
\label{eq:tau_hat_adj}
\hat{\tau}_{\adj} = \sum_{t=1}^T \frac{n_t}{n}\hat{\tau}_{t,\adj}
\end{equation}
However, the adjusted within-block estimators $\hat{\tau}_{t,\adj}$ are not simple difference-in-means estimates like $\hat{\tau}_t$ but rather adjusted for several observed pre-treatment covariates --- number of dependents, reported biological sex, age, and base wage --- using the interacted linear regression method of Lin (2013), as implemented by the \texttt{estimatr} package. We initially tried to additionally adjust for racial identity and educational attainment (which were also available to us), but the presence of sparse levels of these factors in some weeks made covariance matrix estimates singular and therefore prevented the computation of confidence intervals.

In more detail, each adjusted within-block estimator $\hat{\tau}_{t,\adj}$ is computed by mean-centering each covariate and then taking the coefficient of the treatment indicator in a linear regression (fit using ordinary least squares to the observations from week $t$) to predict the desired outcome from the treatment indicator, the mean-centered covariates, and all of the two-way interactions between the treatment indicator and each of the mean-centered covariates. The final aggregated estimate $\hat{\tau}_{\adj}$ is then provably asymptotically unbiased for the sample average treatment effect and cannot have higher asymptotic variance than the non-adjusted estimator $\hat{\tau}$, as shown by Lin (2013). Asymptotically valid confidence intervals and $p$-values can be constructed using heteroskedasticity-robust standard error estimates from the linear regressions; we use the “HC2” estimator of Mackinnon and White (1985). Specifically, we obtain the HC2 standard error estimates $\widehat{se}(\hat{\tau}_{t,\adj})$ for each of the covariate-adjusted within-block average treatment effect estimates, and then let
\[
\widehat{se}(\hat{\tau}_{\adj}) = \sum_{t=1}^T \left(\frac{n_t}{n} \cdot \widehat{se}(\hat{\tau}_{t,\adj})\right)^2
\]
be the standard error estimate of the final estimator $\hat{\tau}_{\adj}$.

We also estimate the average reduction in collected UI benefits among those selected for RESEA using~\eqref{eq:tau_hat_adj} with the outcome $Y$ being the product of the weekly UI benefit rate and the number of weeks spent on UI in the benefit year of the claim, except with the block weights $n_t/n$ modified to be proportional to the number of individuals \emph{selected} for RESEA in each week $t$, given by $\sum_{i=1}^{n_t}W_{ti}$. This is to estimate an “average treatment effect on the treated” instead of an overall average treatment effect.

\subsection{Treatment effect heterogeneity analysis methodology}
We used causal forests (Wager \& Athey, 2018) to estimate the conditional average treatment effect (CATE) of RESEA selection on our three primary outcomes of interest, enabling us to determine subgroups of the population that we expect to benefit the most from RESEA. Causal forests are a variant of the popular random forest machine learning algorithm for prediction (Breiman, 2001) designed to target high precision in estimating the CATE. The CATE refers to the average treatment effect for the subgroup of individuals defined by particular values of one or more available covariates; we notate the CATE by the function $\tau(X)$ where $X$ is a vector of covariates. The CATE predictions from a causal forest are simple averages of CATE predictions from a large number of causal decision trees. Each causal decision tree is trained using a recursive binary splitting algorithm, similar to the standard method to train standard decision trees for predictions, except that the splits are designed to split the training observations into groups with treatment effect predictions that are as different as possible (rather than raw outcomes that are as different as possible). For our analysis, we considered the following covariates, all available in the claim data provided by RIDLT: the number of dependents, reported biological sex, racial identity, age, educational attainment, and base wage. To respect the blocked randomization of our experiment, for each primary outcome we fit a separate causal forest to the claimants randomized in each week of the study period. The fitting for each block is performed using the \texttt{causal\_forest()} function in the \texttt{grf()} package in R (Tibshirani et al., 2024) with all categorical covariates represented using one-hot encoding, the default hyperparameters, and the propensity score set to be constant and equal to the proportion of treated subjects in that block. We then use these causal forests to construct final CATE estimates at each of the covariate vectors $X$ observed in the study cohort as follows. Let $t$ be the week in which an individual with covariate vector $X$ appeared. For each week $t’ = 1,\ldots,T$ of the study period we have a trained causal forest that gives a prediction $\hat{\tau}_{t'}(X)$ of the treatment effect on the outcome of interest for that individual. When $t'=t$, we use an out-of-bag prediction to prevent overfitting bias. Then we compute the predicted CATE $\hat{\tau}(X)$ by taking a weighted average of all $T$ predictions, with weights proportional to the number of claimants in each week. This mirrors our average effect analysis:
\[
\hat{\tau}(X) = \sum_{t=1}^T \frac{n_t}{n} \hat{\tau}_t(X)
\]
If the true underlying CATE function varies with week, as we might expect due to unobserved differences in the study population across weeks, we can interpret $\hat{\tau}(X)$ as the estimated effect of RESEA selection on an individual with covariate vector $X$ provided they were sampled uniformly from the population defined by the aggregation of all of our study subjects across all weeks.

\section{Results}

\begin{table}[!tbp]
\centering
\renewcommand{\arraystretch}{1.8}
\begin{tabular}{|c|c|c|}
\hline
& \textbf{Treatment group (n = 11,700)} & \textbf{Control group (n = 11,849)} \\
\hline
Proportion female & 50.1\% & 48.3\% \\
\hline
Mean base wage & \$49,175 & \$49,748 \\
\hline
Mean age & 45.2 years & 45.2 years \\
\hline
Mean number of & 0.44 & 0.43 \\
dependents & & \\
\hline
Proportion white alone & 63.2\% & 64.7\% \\
\hline
Education & 3.9\% no high school degree & 4.5\% no high school degree \\
& 19.3\% high school diploma & 20.8\% high school degree \\
& 21.4\% some college or Associates & 21.5\% some college or Associates \\
& 8.8\% Bachelors & 8.9\% Bachelors \\
& 3.8\% Graduate & 3.8\% Graduate \\
& 42.7\% missing (phone filer) & 40.5\% missing (phone filer) \\
\hline
\end{tabular}
\vspace{0.2cm}
\caption{Unweighted pre-treatment characteristics of the aggregate treatment and control groups in the study cohort. Note that education information was not available for those who filed their UI claims by phone.}
\end{table}

Basic pre-treatment demographic characteristics of the aggregate treatment and control groups are provided in Table 1. The groups were broadly similar, but we note that this does not necessarily indicate, in itself, that randomization was successful. Since the randomized treatment assignment was performed on a weekly basis, and the number of UI applicants varied substantially across weeks with the number of treated claimants generally fixed across weeks for staffing reasons, the aggregate treatment and control groups differ in the relative weights of the populations of UI claimants across different weeks. Evidence for the success of the randomization is provided through the weekly balance tests pictured in Fig. 1, as discussed previously.

\begin{table}[htbp]
\centering
\renewcommand{\arraystretch}{1.2}
\begin{tabular}{|c|c|c|c|c|}
\hline
\textbf{Outcome} & \textbf{Treatment} & \textbf{Control} & \textbf{Treatment-control} & \textbf{Average treatment} \\
& & & \textbf{difference} & \textbf{effect estimate} \\
\hline
Wage & \$33,073 & \$33,043 & \$29 & \$1,153 \\
& (SD: 36,749) & (SD: 36,231) & $p = 0.95$ & $p = 0.022$ \\
& & & 95\% CI: [-903, 962] & 95\% CI: [166, 2,141] \\
\hline
Reemployment & 75.2\% & 75.0\% & 0.26 pp & 1.47 pp \\
& & & $p = 0.64$ & $p = 0.013$ \\
& & & 95\% CI: [-0.84, 1.37] & 95\% CI: [0.32, 2.62] \\
\hline
Weeks on UI & 15.2 weeks & 17.3 weeks & -2.10 weeks & -1.99 weeks \\
& (SD: 9.3) & (SD: 9.4) & $p < 0.001$ & $p < 0.001$ \\
& & & 95\% CI: [-2.34, -1.86] & 95\% CI: [-2.24, -1.75] \\
\hline
Short-term & \$23,776 & \$23,897 & -\$121 & \$1,195 \\
wage & (SD: 31,948) & (SD: 32,293) & $p = 0.77$ & $p = 0.007$ \\
& & & 95\% CI: [-941, 700] & 95\% CI: [326, 2,064] \\
\hline
Short-term & 61.4\% & 61.6\% & -0.18 pp & 1.66 pp \\
reemployment & & & $p = 0.77$ & $p = 0.012$ \\
& & & 95\% CI: [-1.43, 1.06] & 95\% CI: [0.37, 2.96] \\
\hline
Short-term & 14.2 weeks & 15.9 weeks & -1.70 weeks & -1.77 weeks \\
weeks on UI & (SD: 8.4) & (SD: 8.3) & $p < 0.001$ & $p < 0.001$ \\
& & & 95\% CI: [-1.92, -1.49] & 95\% CI: [-1.99, -1.55] \\
\hline
\end{tabular}
\vspace{0.2cm}
\caption{The average values of our long-term and short-term outcomes for the aggregate treatment and control groups in the study cohort, along with information about the differences in these averages between the treatment and control groups and our average treatment effect estimates. For the average outcomes, standard deviations are given in parentheses for the quantitative outcomes (wage and weeks on UI). For the treatment-control differences and the average treatment effect estimates, we provide two-sided $p$-values and 95\% confidence intervals based on Welch's two-sample $t$-test and the methodology described in the main text, respectively. The two rightmost columns illustrate how direct comparisons of outcome averages between the aggregate treatment and control groups can lead to significantly biased causal conclusions.}
\end{table}

Overall, we find that RESEA selection led to a statistically significant improvement in all three of the preregistered outcomes. RESEA selection led to a wage outcome that was \$1,153 higher annually (95\% CI: [\$166, \$2,141]; $p = 0.022$), a 3.5\% increase relative to the mean wage outcome of \$33,043 in the control group (Table 2). There was also an estimated 1.47 percentage point increase in reemployment (95\% CI: [0.32, 2.62]; $p = 0.013$) and a 1.99 week reduction in the number of weeks on UI (95\% CI: [1.75 weeks, 2.24 weeks]; $p < 0.001$) due to RESEA selection. The latter is an 11.5\% decrease compared to the mean control group weeks on UI outcome of 17.3 weeks. All three of these results remain statistically significant after correcting for multiple testing using Holm’s method to keep the familywise error rate below 5 percent.

The results for the short-term outcomes are substantively similar to those for the longer-term outcomes (Table 2). We estimate RESEA selection leads to a short-term wage outcome that is \$1,195 higher (95\% CI: [\$326, \$2,064]; $p$ = 0.007) on an annualized basis and a 1.66 percentage point higher rate of short-term reemployment (95\% CI: [0.37, 2.96]; $p$ = 0.012). Given that the magnitudes of these causal estimates are very similar to those for the corresponding long-term outcomes, this suggests that the full impact of RESEA selection on wages and reemployment appears quickly by the second quarter after randomization, and persists for at least five quarters. We also find that those selected for RESEA spend 1.77 fewer weeks on UI in the benefit year of their claim (95\% CI: [1.55 weeks, 1.99 weeks]; $p$ < 0.001). As this is a bit smaller than the effect size of 1.99 weeks estimated for the long term weeks on UI outcome, which includes an additional 26 weeks beyond the benefit year, we see some directional evidence that RESEA selection may also reduce the time spent on UI in the following year, suggesting that it may be helping claimants find employment that lasts longer, thereby reducing repetitive use of UI benefits. 

As noted earlier, wages and reemployment are artificially zero-inflated due to missing out-of-state wage data. The Rhode Island job market is tightly integrated with those of the neighboring states of Massachusetts and Connecticut; the 2016-2020 5-Year American Community Survey found that nearly 16 percent of Rhode Island workers 16 years and older commuted out of state, mainly to Massachusetts. We note, however, that this missingness should not affect the directional validity of our causal analysis. This is because randomization ensures that this missingness is balanced between the treatment and control groups (within each block), unless RESEA selection itself differentially helps or hinders finding employment out of state or self-employment. To investigate this, we examined the presence of individuals who failed to exhaust their UI benefits in the benefit year of their claim while having \$0 in reported RI wages for all of the five complete quarters after randomization. There were 3,325 claimants falling into this group, including 1,618 individuals (13.7\%) in the aggregate control group and 1,707 individuals (14.6\%) in the aggregate treatment group. After controlling for randomization week, this difference in proportions was not significant ($p$ = 0.55). The lack of statistical significance on this test with our large sample size along with the fact that the Rhode Island job database focuses overwhelmingly on jobs offered by Rhode Island employers suggests that RESEA selection does not differentially assist in finding out-of-state employment or self-employment. 

\subsection{Cost-benefit analysis}
We find that the decrease in time spent drawing UI benefits due to RESEA selection corresponds to a considerable cost savings to the state from paying out a smaller amount of UI benefits. We estimate that the average individual selected for RESEA during the study period drew \$676 less (95\% CI: [\$563, \$789], $p < 0.001$) in UI benefits during the same benefit year of their claim than they would have if they were not selected for RESEA. This is a 9.9 percent reduction relative to the control group average of \$6,839 in UI benefits drawn during the benefit year of the claim. Multiplying the estimate by the total number of individuals selected for RESEA in the study ($n = 11,700$) suggests that as a result of the program, the state paid out about \$7.9 million fewer in UI benefits to claimants randomized during the study period over the same benefit year as their UI claims. This is much higher than the total expenditures of \$2.99 million for the RESEA program over the study period, representing \$2.64 in savings for the state budget for every \$1 spent on the program.

\subsection{Heterogeneous effects analysis}
\begin{figure}[!tbp]
\centering
\includegraphics[width=\linewidth]{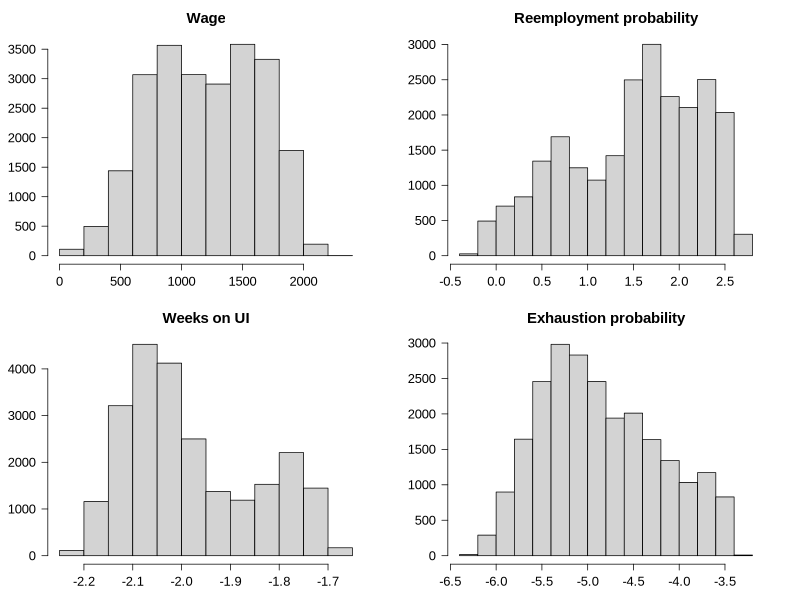}
\caption{ Histograms of the CATE estimates across the study cohort for the wage outcome (top left), reemployment outcome (top right, in percentage points), weeks on UI outcome (bottom left), and exhaustion probability (bottom right, in percentage points). The CATE estimates are weighted averages of individual-level CATE estimates from a series of causal forests trained using the grf package in R (Tibshirani et al., 2025). Each causal forest is trained on the data from within a particular randomization week (block), and the weights are proportional to the block sizes.}
\label{fig:cate_histograms}
\end{figure}

\begin{figure}[!tbp]
\centering
\includegraphics[width=\linewidth]{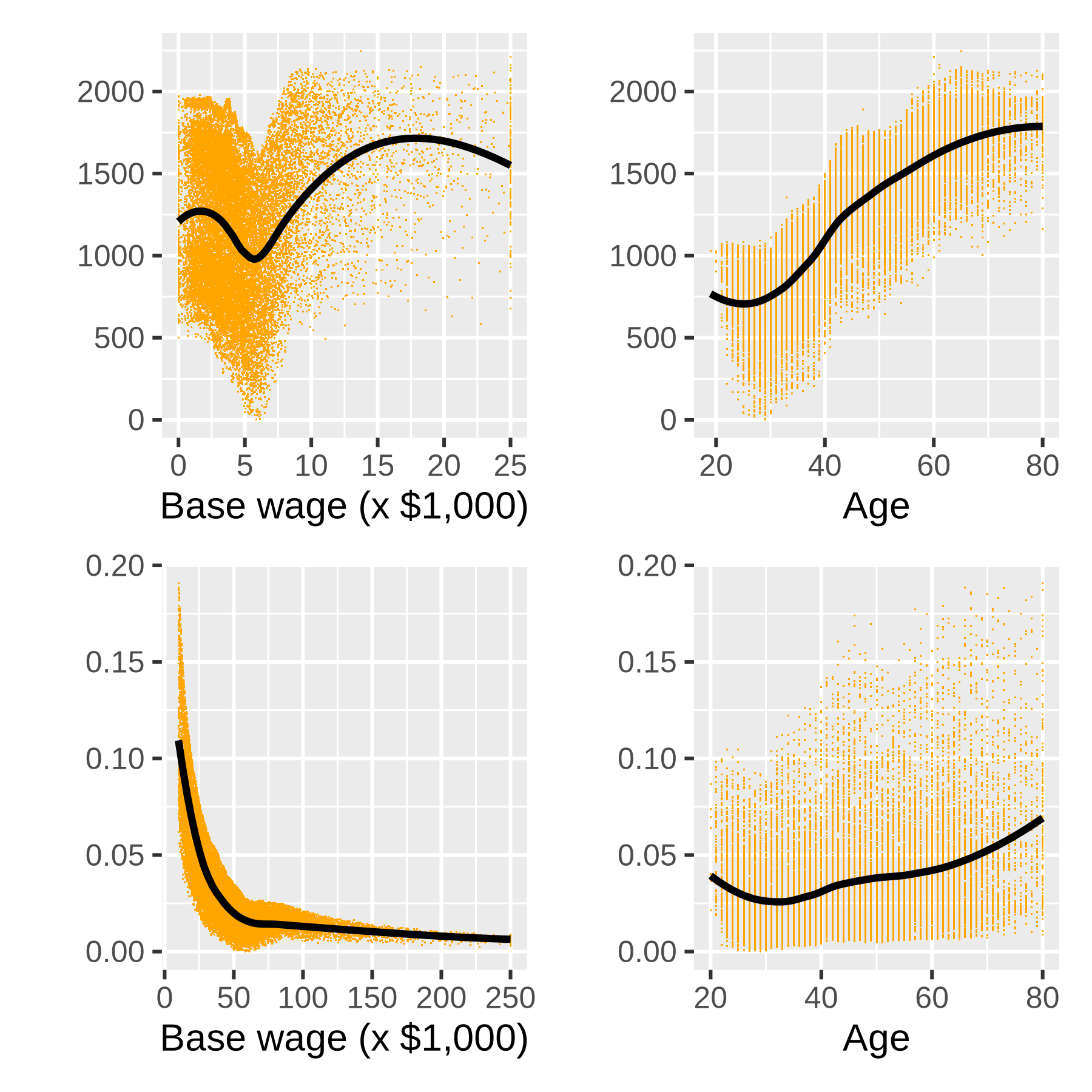}
\caption{The estimated CATE for the wage outcome for all individuals in the study cohort versus their base wage (top left) and age (top right). The bottom plots show the same information but with the vertical axis a fraction of base wage. The black lines are scatterplot smoothers from locally weighted regression computed with the loess() function in R, with default smoothing parameter span. Base wage (resp. age) is truncated to \$250,000 (resp. 80 years) for all plots. In the bottom row plots, individuals with base wage less than \$10,000 are also excluded. 
}
\label{fig:wage_cate_plot}
\end{figure}

\begin{figure}[!tbp]
\centering
\includegraphics[width=0.8\linewidth]{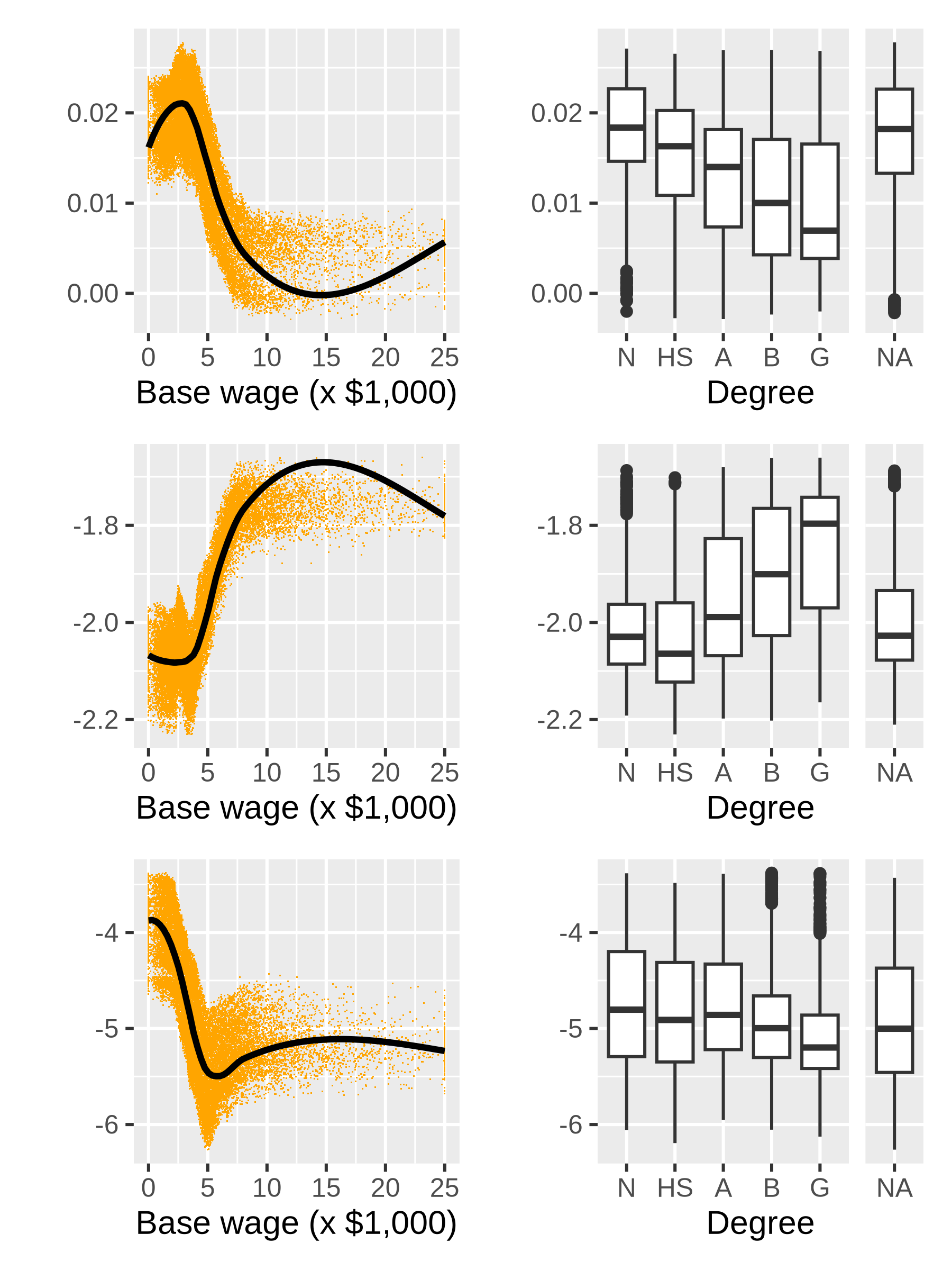}
\caption{The estimated CATE for the reemployment outcome (top row, in percentage points), the weeks on UI outcome (middle row), and the chance of exhaustion (bottom row, in percentage points) for all individuals in the study cohort versus their base wage (left column, truncated at \$250,000) and highest degree obtained (right column). Degree abbreviations are as in Fig. 3.}
\label{fig:non_wage_cate_plot}
\end{figure}

CATE estimates for the wage outcome, computed using causal forests as described in the Methods section, are positive for all individuals in the study (Fig. 4). The estimated wage effect is largest for older workers, even when computing these effects as a fraction of their typically higher base wages (Fig. 5). Higher-earning and better educated workers have notably smaller estimated reemployment effects than average. However, this same group is still estimated to spend nearly 2 fewer weeks on UI on average due to RESEA selection. This is driven by a larger decrease in UI exhaustion rates (e.g. using the maximum of 26 weeks of UI benefits during the benefit year) than their lower-earning and less educated counterparts (Fig. 6). Thus, while RESEA selection may not be strongly nudging such claimants to find employment, it is evidently still encouraging them to draw fewer weeks of UI benefits, translating to program cost savings.

\section{Discussion}
Research shows that employment provides meaning and yields beneficial health and social outcomes, making it in the state's interest that individuals return to work as soon as possible. At the same time, the government must ensure that unemployment insurance funds remain sufficient to cover everyone who needs them. By helping unemployment insurance recipients return to work more quickly, programs like RESEA are designed to benefit individual job seekers while preserving resources for future claimants. Our findings provide strong causal evidence that the RESEA program in Rhode Island is effective at achieving both these goals.
In particular, our findings present strong causal evidence that selection into Rhode Island’s RESEA program had a beneficial impact on wages, reemployment, and duration of UI benefits, the latter of which translates into a large cost savings for the government even after accounting for program expenditures. Consistent with existing literature, the effect on UI duration appears to be the most salient of the three outcomes (e.g., Klerman et al., 2019).

The magnitude of the wage effect can plausibly be explained largely by an earlier return to work as implied by the effect on weeks on UI. As a heuristic calculation, we note that spending 1.99 fewer weeks on UI and instead earning the average control group annual wage of \$33,043 corresponds to a \$1,265 increase in annual wages assuming 52 weeks in a year. This is quite close to the wage effect estimate of \$1,153 and consistent with our CATE estimates suggesting that RESEA selection causes a slightly greater reduction in time spent on UI for individuals with lower base wages. However, this is not the only plausible explanation for how RESEA selection causes higher wages. Finally, we note that the magnitudes of our estimated short-term and long-term effects are similar, suggesting that most of the effects of RESEA show up quickly—within two quarters of randomization—and sustain through the long term.

\subsection{Methodological lessons}

\begin{figure}[!tbp]
\centering
\includegraphics[width=\linewidth]{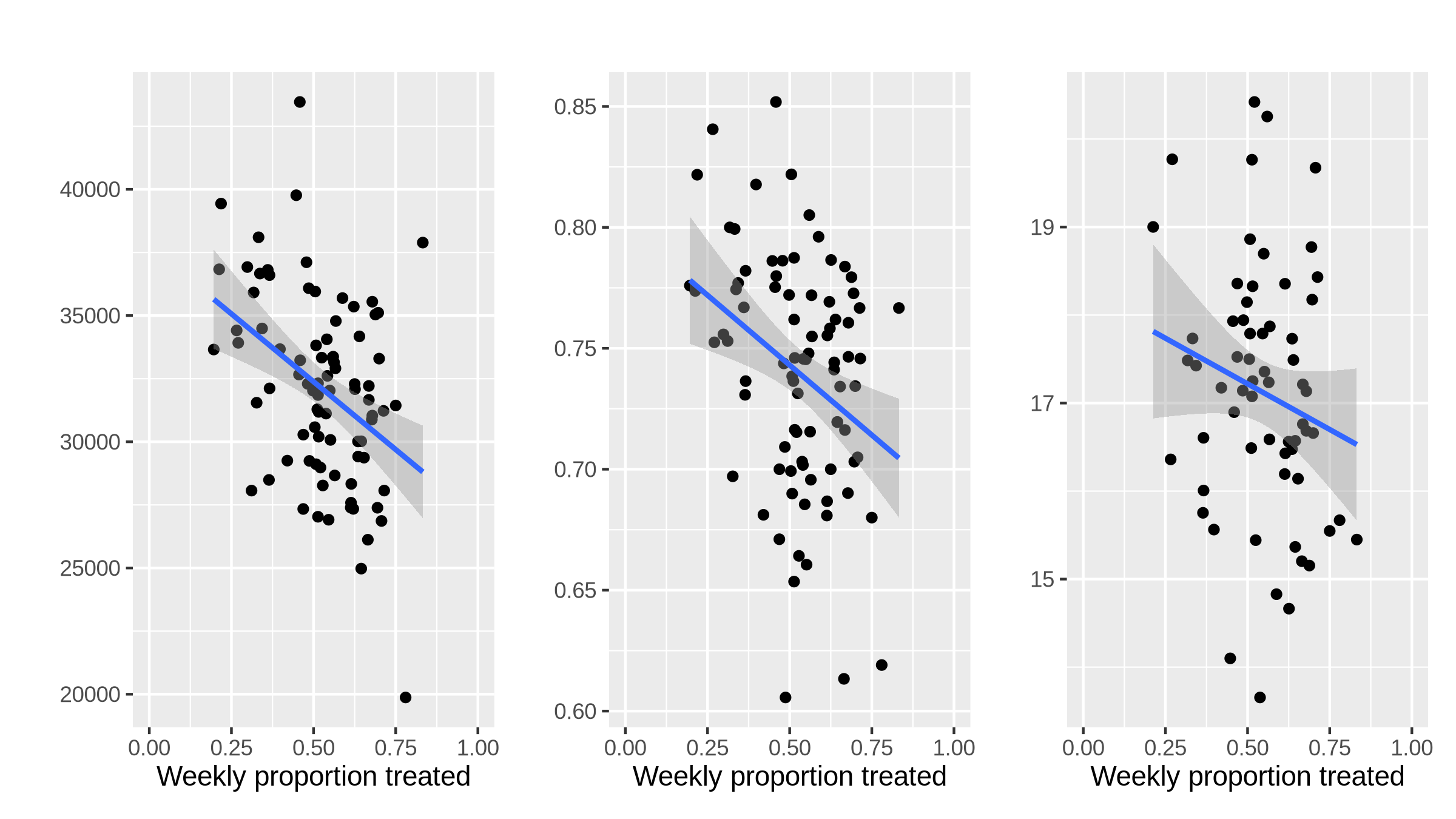}
\caption{Scatterplots showing the average wage outcome (left), the average reemployment outcome (middle), and the average weeks on UI outcome (right) in the control group for each week (block) during the study period versus the fraction of individuals assigned to the treatment group in that block. Weeks with a higher proportion of treated claimants tended to have significantly smaller outcomes, illustrating that randomization week is a substantial confounder for all three outcomes. The blue lines are each computed by fitting an ordinary least squares regression to the points, with the surrounding shaded region indicating a 95\% confidence band.}
\label{fig:confounding}
\end{figure}

\begin{table}[!tbp]
\centering
\renewcommand{\arraystretch}{1.5}
\begin{tabular}{|c|c|c|c|}
\hline
\textbf{Outcome} & \textbf{Difference in means} & \textbf{Covariate adjusted} & \textbf{Fixed effects} \\
& & & \textbf{regression} \\
\hline
Wage & \$1,153 & \$1,137 & \$1,100 \\
& $p = 0.022$ & $p = 0.014$ & $p = 0.027$ \\
& 95\% CI: [166, 2,141] & 95\% CI: [228, 2,047] & 95\% CI: [127, 2,073] \\
\hline
Reemployment & 1.47 pp & 1.48 pp & 1.49 pp \\
& $p = 0.013$ & $p = 0.012$ & $p = 0.011$ \\
& 95\% CI: [0.32, 2.62] & 95\% CI: [0.32, 2.63] & 95\% CI: [0.34, 2.64] \\
\hline
Weeks on UI & -1.99 wks & -1.93 wks & -1.97 wks \\
& $p < 0.001$ & $p < 0.001$ & $p < 0.001$ \\
& 95\% CI: [-2.24, -1.75] & 95\% CI: [-2.18, -1.68] & 95\% CI: [-2.22, -1.72] \\
\hline
\end{tabular}
\vspace{0.2cm}
\caption{A comparison of the ITT estimates, two-sided $p$-values, and 95\% confidence intervals for our primary outcomes from three different methodologies: the preregistered analysis based on eq. (1), the covariate adjusted analysis based on eq. (2), and the fixed effects regression in eq. (3). The leftmost column based on the preregistered analysis duplicates the rightmost column in Table 2 for ease of comparison.}
\end{table}

If we directly compared treatment group and control group outcome averages without accounting for the week of randomization, the results would suggest RESEA selection had no detectable average effect on the wage and reemployment outcomes (column 4, Table 2). However, such an analysis is potentially biased under the experimental design because it does not respect the blocked randomization, which makes the week of randomization a known confounder. For example, weeks with a larger number of UI applicants, which had a higher proportion of control group individuals, tended to have lower-earning applicants (Fig. 7), thereby inflating the average wage outcome in the control group relative to the treatment group. This is critical because out of the thirteen publically available studies of RESEA-related employment assistance programs rated by CLEAR as having high causal evidence, six used block randomization like we do, yet only two of these (Black, Smith, Berger, \& Noel., 2002; Klerman et al., 2019) properly adjusted for the block. Three others among these six studies used regression or matching methods in an attempt to control for observed demographic variables but not the block (Poe-Yamagata et al., 2011; Michaelides et al., 2012; Manoli et al., 2018), while the remaining study (Michaelides \& Mueser, 2018) fit a linear regression model with the block indicators as fixed effects, as did two other more recent evaluations in Nevada (Michaelides \& Mian, 2021; Brigandi et al., 2024) . These fixed effects regressions amount to estimating the coefficient  in the regression model
\begin{equation}
\label{eq:fixef}
Y_{ti} = \alpha_t + \tau W_{ti} + \epsilon_{ti}
\end{equation}
where $\alpha_t$ is a non-random intercept specific to the randomization week (block). However, in general, since the proportion of individuals treated varied across blocks, the fixed effects regression estimator is only unbiased if the true treatment effect is constant across blocks. Otherwise, the fixed effects estimator will estimate a weighted average treatment effect, with subjects in different blocks receiving different weights based on the proportion of treated subjects in their blocks, which is much more difficult to interpret (Angrist, 1998). This issue is discussed in more detail by Gibbons et al. (2019). Table 3 compares the results reported using our methodology with those from a fixed effects regression model.

Consequently, we recommend a re-analysis of the data from the prior studies that used block randomization but did not adjust for the block as a known confounder. In the absence of pre-treatment covariates, the block size-weighted average of difference-in-means estimates works well as it is unbiased for the average treatment effect (see eq. (1)). If baseline covariates are observed, we would recommend instead using the block size-weighted average of linearly adjusted within-block estimates, as in eq. (2). Assuming the number of covariates adjusted for is small relative to the number of observations in each block, this cannot increase the asymptotic variance while retaining asymptotic unbiasedness as shown by Lin (2013), and thus will lead to more precise estimates and narrower confidence intervals. We hope this observation is useful for future pre-registered block randomized program evaluations, whether of RESEA or other related or unrelated programs.

\section*{Appendix}

\setcounter{table}{0}
\renewcommand{\tablename}{Appendix Table}
\begin{longtable}{| >{\raggedright\arraybackslash}p{0.13\textwidth} | >{\raggedright\arraybackslash}p{0.10\textwidth} | >{\raggedright\arraybackslash}p{0.13\textwidth} | >{\raggedright\arraybackslash}p{0.15\textwidth} | >{\raggedright\arraybackslash}p{0.15\textwidth} | >{\raggedright\arraybackslash}p{0.27\textwidth} |}
\caption{Literature review summary of the studies in the DOL CLEAR database rated as having “high causal evidence” for evaluating RESEA (although none of the studies directly study an RESEA program). Rows highlighted in blue correspond to studies that do not appear to be readily available on the Internet.} \label{tab:studies} \\
\hline
\textbf{Reference} & \textbf{Study period} & \textbf{Location(s)} & \textbf{Design} & \textbf{Properly controls for block, if relevant?} & \textbf{Summary of findings} \\
\hline
\endfirsthead

\multicolumn{6}{c}%
{{\bfseries \tablename\ \thetable{} -- continued from previous page}} \\
\hline
\textbf{Reference} & \textbf{Study period} & \textbf{Location(s)} & \textbf{Design} & \textbf{Properly controls for block, if relevant?} & \textbf{Summary of findings} \\
\hline
\endhead

\hline \multicolumn{6}{|r|}{{Continued on next page}} \\ \hline
\endfoot

\hline
\endlastfoot

Manoli et al. (2018) & Q3-Q4 2009 & NV & Block-randomized with unequal treatment fractions across blocks & No: Uses propensity score matching on demographic variables, discards substantial proportion of observations due to poor overlap & Studied REA in NV. \par\medskip Treatment resulted in higher employment and earnings relative to the control group for each of the six years following the intervention. \par\medskip Treatment resulted in fewer weeks on UI for the first year after intervention. \\ \hline

Klerman et al. (2019) & 2015-2016 & IN, NY, WA, WI & Block-randomized with unequal treatment fractions across blocks (IN). \par\medskip Block-randomized with (nearly) equal treatment fractions across blocks (NY, WA, WI) & Yes (IN): Observations are reweighted based on treatment fraction in a fixed regression. \par\medskip N/A (other states): Block is not a confounder, and a fixed effects regression is used. & Studied REA in four states. \par\medskip Participants in the multiple REA treatment group spent fewer weeks on UI than the single REA treatment group. \\ \hline

\rowcolor{blue!15} Behrens (1987) & March 1985 - May 1986 & Hackensack, NJ & & & \\ \hline

Decker et al. (2000) & 1995-1996 & Washington, DC and FL & Simple random assignment & N/A: No blocks & Studied the Job Search Assistance Demonstration program. \par\medskip Treatment reduced UI benefits and increased earnings compared to control group in Washington, D.C., but not in Florida. \\ \hline

\rowcolor{blue!15} Corson et al. (1985) & 1983 & Charleston, SC & & & \\ \hline

Black et al. (2003) & 1994-1996 & KY & Blocked tie-breaker design based on predicted probability of benefit exhaustion, with unequal treatment fractions across blocks & Yes: Uses fixed effects estimator but careful to note interpretation of estimands due to unequal treatment fractions & Studied the Worker Profiling and Reemployment Services program. \par\medskip Participants in the treatment group had a statistically significant reduction in weeks on UI (2.2 weeks) during the six-quarter follow-up period, compared to the control group but no impact on UI exhaustion or total UI benefits received. \\ \hline

\rowcolor{blue!15} ERP project final report & 1984 & WI & & & \\ \hline

Michaelides et al. (2012) & Q3-Q4 2009 & NV & Block-randomized with unequal treatment fractions across blocks & No: Uses a fixed effects regression that only adjusts for demographic and employment variables & Studied REA in NV. \par\medskip Intervention resulted in a reduction in the average duration and amount of UI benefits receipt, an increase in employment rates, and an increase in earnings compared to control group. \\ \hline

Corson et al. (1989) & 1986-1987 & NJ & Block-randomized with (nearly) equal treatment fractions across blocks & N/A: Block is not a confounder, reports regression adjusted differences & Studied the New Jersey Unemployment Reemployment Demonstration Project. \par\medskip Main results are based on surveys, which are argued to be more reliable than wage records. No significant result found in wages. \\ \hline

Anderson et al. (1991) & 1986-1987 & NJ & Block-randomized with (nearly) equal treatment fractions across blocks & N/A: Block is not a confounder, reports regression adjusted differences & Studied the New Jersey Unemployment Reemployment Demonstration Project. \par\medskip Treatment resulted in fewer UI dollars received (\$293), and fewer weeks spent on UI (1.6 weeks) compared to control group. \par\medskip No effects for reemployment, earnings, or weeks worked. \\ \hline

Corson and Haimson (1996) & 1986-1987 & NJ & Block-randomized with (nearly) equal treatment fractions across blocks & N/A: Block is not a confounder, reports regression adjusted differences & Studied the New Jersey Unemployment Reemployment Demonstration Project. \par\medskip Treatment resulted in a statistically significant reduction in UI dollars received and weeks spent on UI, compared with the control group. \par\medskip No effects were found on the probability of working, level of earnings, or weeks worked. \\ \hline

Benus et al. (2008) & 2005-2006 & ND & Simple randomized design (based on last digit of SSN) & N/A: No blocks & Studied REA in North Dakota. \par\medskip No statistically significant impacts of REA on UI benefits receipt, employment, or earnings were found as a result of REA treatment. \\ \hline

Poe-Yamagata et al. (2011) & & FL, ID, IL, NV & Block-randomized with unequal treatment fraction across blocks (FL, NV). \par\medskip Block-randomized (ID, IL) with unclear treatment fractions. & No: Uses a regression analysis that does not control for randomization week & Studied REA in four states. \par\medskip Treatment resulted in statistically significant reductions in weeks spent on UI, total amount of UI benefits received, and the probability of benefit exhaustion. Treatment also increased probability of employment and earnings over the four follow-up quarters. \\ \hline

Lachowska et al. (2015) & 1986-1987 & WA & Simple randomized design (based on last digit of SSN) & N/A: No blocks & Studied effects of eliminating work-search requirements for collecting UI benefits. \par\medskip The treatment group with less-stringent work-search requirements was significantly less likely to be employed in the first quarter following their claims, compared with the groups with more-stringent requirements. The group with less stringent work search requirements also received more UI benefit payments for more weeks and exhausted UI benefits at a higher rate during the year following their initial claims. \\ \hline

Lachowska et al. (2016) & 1986-1987 & WA & Simple randomized design (based on last digit of SSN) & N/A: No blocks & Studied effects of eliminating a work test as a requirement for collecting UI benefits. \par\medskip Extended Lachowska (2015) and found that the treatment groups with more-stringent work search requirements were more likely to be employed compared to the less-stringent requirement group in the first year following their claims. More stringent work requirement groups also received UI benefits payments for fewer weeks, exhausted UI benefits at a lower rate, and received fewer conditional payments in the year following their initial UI claim. \\ \hline

Michaelides and Mueser (2018) & Q3-Q4 2009 & NV & Block-randomized with unequal treatment fractions across blocks & Only if treatment effect constant across blocks; uses fixed effects regression & Studied REA in NV. \par\medskip Studied a treatment group who received job matching and work search preparation services, a treatment group with only REA eligibility services and job search activity training, services, and a control group that was only asked to track their job search. Found that the first treatment group had significantly higher reemployment rates and lower UI benefit receipt but similar earnings as compared to the control group. \\

\end{longtable}

\begin{longtable}{| >{\raggedright\arraybackslash}p{0.13\textwidth} | >{\raggedright\arraybackslash}p{0.10\textwidth} | >{\raggedright\arraybackslash}p{0.13\textwidth} | >{\raggedright\arraybackslash}p{0.15\textwidth} | >{\raggedright\arraybackslash}p{0.15\textwidth} | >{\raggedright\arraybackslash}p{0.27\textwidth} |}
\caption{Same as Appendix Table 1, but for two recent RESEA evaluation reports that have not been reviewed by USDOL as of the time of writing.} \label{tab:resea_evals} \\
\hline
\textbf{Reference} & \textbf{Study period} & \textbf{Location(s)} & \textbf{Design} & \textbf{Properly controls for block, if relevant?} & \textbf{Summary of findings} \\
\hline
\endfirsthead

\multicolumn{6}{c}%
{{\bfseries \tablename\ \thetable{} -- continued from previous page}} \\
\hline
\textbf{Reference} & \textbf{Study period} & \textbf{Location(s)} & \textbf{Design} & \textbf{Properly controls for block, if relevant?} & \textbf{Summary of findings} \\
\hline
\endhead

\hline \multicolumn{6}{|r|}{{Continued on next page}} \\ \hline
\endfoot

\hline
\endlastfoot

Brigandi et al. (2024) & Dec. 2021 - Dec. 2022 & WA & Block-randomized with unequal treatment fractions across blocks & Only if treatment effect constant across blocks; uses fixed effects regression & Studied RESEA in WA. \par\medskip Study evaluated if a scheduling process change would reduce the number of benefit disqualifications. Treatment was found to reduce the number of no-shows and the number of benefit disqualifications that resulted from not making/attending an appointment. The UI claimants who benefitted most were more likely to self-identify as African American/Black and male, were younger, and had lower earnings than those who benefited less. \\ \hline

Michaelides and Mian (2021) & 2014-15 & NV & Block-randomized with unequal treatment fractions across blocks & Only if treatment effect constant across blocks; uses fixed effects regression & Studied REA in NV. \par\medskip Found that treatment resulted in an increase in earnings for treated claimants over study follow-up periods ranging from 1.5 to 5 years after random assignment and reduced UI payments. \\

\end{longtable}

\end{document}